\documentclass[reprint,aps,secnumarabic,balancelastpage,amsmath,amssymb,floatfix,superscriptaddress,nofootinbib]{revtex4-1}
\usepackage[latin1]{inputenc}
\usepackage{amsmath}
\usepackage{amsfonts}
\usepackage{bbold}
\usepackage{amssymb}
\usepackage[pdftex]{graphicx}

\usepackage{hyperref}
\hypersetup{
	colorlinks = true,
	citecolor = {red}
	}

\let\vec\mathbf

\begin{document}

\title{\bf High-Field Expansion Approach to Kagome Antiferromagnets with Dzyaloshinskii-Moriya Interactions}

\author{Michael O. Flynn}
\email{miflynn@ucdavis.edu}
\author{Rajiv R. P. Singh}
\email{singh@physics.ucdavis.edu}
\affiliation{Department of Physics, University of California Davis, CA, 95616, USA}

\date{\today}

\begin{abstract}

We apply linked cluster expansion techniques to study the polarized high-field phase of a spin-half antiferromagnet on the Kagome lattice with Heisenberg and Dzyaloshinskii-Moriya interactions (DMI). We find that the Dirac points of the single-magnon spectrum without DMI are robust against arbitrary DMI when the magnetic field lies in the Kagome plane. Unlike the typical case where DMI gaps the spectrum, here we find that varying the DMI merely shifts the location of the Dirac points. In contrast, a magnetic field with a component out of the Kagome plane gaps the spectrum, leading to topological magnon bands. We map out a topological phase diagram as the couplings are varied by computing the band Chern numbers. A pair of phase transitions are observed and we find an enhanced thermal Hall conductivity near the phase boundary.

\end{abstract}

\maketitle

\section{Introduction}

In recent decades, the search for novel phases of matter in frustrated systems has been a major focus of condensed matter physics \cite{Balents2010,Yan1173,2015NatPh..11..444G,PhysRevLett.64.88}. Exotic phases of matter formed due to the interplay of various physical interactions can be delicate, forcing theorists to consider models with a range of perturbations. One such perturbation is Spin-orbit coupling (SOC), which enters the Hamiltonian of insulating magnets via  Dzyaloshinskii-Moriya interactions (DMI) \cite{1958JPCS....4..241D,PhysRev.120.91}.

SOC has recently garnered increased attention \cite{2006AnPhy.321....2K,PhysRevLett.102.017205,2010NatPh...6..376P,2013PhRvB..87f4423L}. Experimental efforts to artificially control SOC are an important aspect of spintronics \cite{Chen2013TailoringTC,article}, and the competition between frustration, SOC, and various symmetry-breaking perturbations has proven to contain rich physics \cite{2017RPPh...80a6502S, 2016NatCo...712691L,2018arXiv181101946O,2018JPCM...30x5803O}. In such systems spin-wave theory has been used to extract the spectrum of quasiparticles (particularly magnons), yielding a host of predictions for insulating ferromagnets and antiferromagnets, with and without DMI. Through these calculations, the theory of magnon transport has been refined and various experimental probes have been proposed.

Despite the successes of spin-wave theory, the approach has some limitations. Linearization of the magnon Hamiltonian forbids certain processes and leaves out terms that may alter magnon spectra and wavefunctions, especially for low spin. It is also important to move beyond single-magnon bands and study the physics of multi-magnon states. Directly addressing spin-half systems would open the possibility of studying multiparticle inelastic scattering processes and bound states in realistic systems.

This study considers a different approach  to the physics of strongly correlated magnetic systems, built from perturbation theory. While perturbation theory has obvious limitations, it removes the linearization inherent in spin-wave theory. The work presented here focuses on single-magnon states, which allows us to make contact with previous studies of similar models. Multi-magnon states will be the subject of future investigations.

In this work, we consider a model of localized spin-$1/2$ particles on the Kagome lattice in high magnetic fields. The model, while simple, furnishes both a magnonic Dirac semimetal and a topological magnon insulator, which can each be achieved by tuning an external magnetic field. Importantly, the Dirac points in the semimetal phase are robust against arbitrary DMI and can be manipulated in principle with modern spintronics techniques. The interactions between the spins are purely nearest-neighbor and include an antiferromagnetic Heisenberg coupling and DMI. The Hamiltonian is

\begin{equation}\label{Ham}
H = -\sum_{i}\vec{B}\cdot \vec{S}_{i}+\sum_{\langle ij\rangle}\left[J\vec{S}_{i}\cdot \vec{S}_{j} + \vec{D}_{ij}\cdot \left(\vec{S}_{i}\times\vec{S}_{j}\right)\right]
\end{equation}

\begin{figure}
  \includegraphics[width=0.75\linewidth]{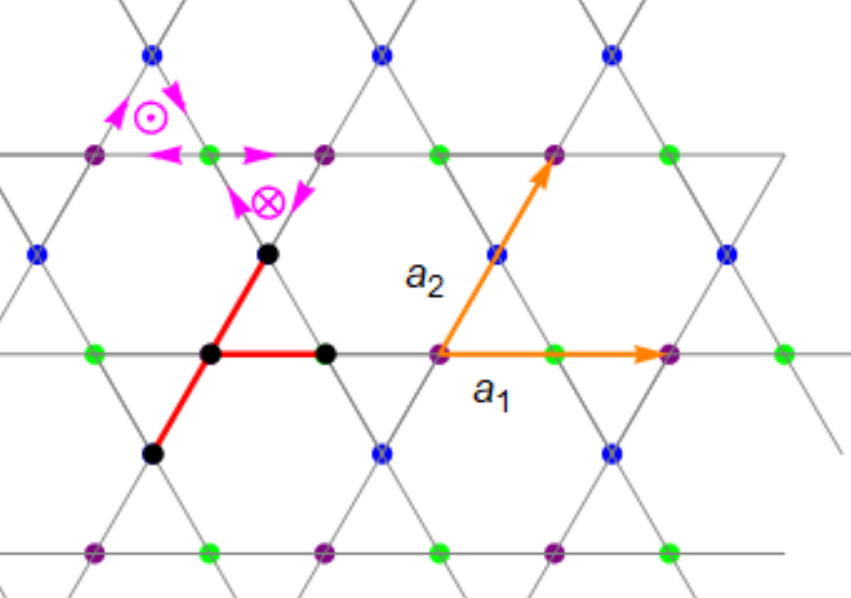}
  \caption{A section of the Kagome lattice. The spins are colored according to their sublattice, and an example of a connected graph with three bonds is highlighted (red). This graph has six (proper, connected) subgraphs. We fix the  lattice spacing so that the Bravais lattice vectors are given by $\vec{a}_{1} = (2,0,0), \vec{a}_{2} = (1,\sqrt{3},0)$ (orange). Our convention for the DMI is shown in purple: cross products are positive along indicated link directions, $D_{z}$ alternates between plaquettes, and $D_{p}$ (not shown) points away from plaquette centers on bonds.}
  \label{fig:graphs}
\end{figure}

Here $\vec{B}$ is the magnetic field, $\langle ij\rangle$ denotes nearest neighbors, $J>0$ is the antiferromagnetic Heisenberg coupling, and $\vec{D}_{ij}$ is the DM vector on the bond $ij$ \cite{PhysRevB.66.014422,PhysRevB.78.140405,2015arXiv150801523S,PhysRevLett.98.207204,PhysRevB.95.054418,PhysRevB.85.104417,PhysRevLett.117.187203,PhysRevB.98.224414,Owerre_2016,PhysRevLett.104.066403,2018PhRvB..97h1106R,2019arXiv190105683C,2019arXiv190504549M}. We will treat the spin interactions as perturbations to the magnetic field coupling; this is valid in the polarized phase. Remaining in the polarized phase roughly requires that $|\vec{B}| \gtrapprox (J+|\vec{D}|)/3$, but more precisely corresponds to magnon bands without a zero-energy mode \citep{PhysRevLett.82.4536}. Figure \ref{fig:graphs} shows our DMI conventions.

Studies of similar models have been conducted in the past, outside of the high-field regime. Mook et. al. \cite{PhysRevB.89.134409} considered a ferromagnetic version of (\ref{Ham}) with next nearest neighbor (NNN) exchange. Laurell and Fiete \cite{2018PhRvB..98i4419L} carried out a spin-wave analysis of the antiferromagnetic version of the same model. The NNN exchange used in these works provides a mechanism for the experimentally observed (weak) dispersion in the bottom band of a Kagome system, which would otherwise be flat \cite{PhysRevB.73.214446,PhysRevLett.96.247201,PhysRevB.92.094409}. Although we do not explicitly include NNN interactions, our approach ultimately produces dispersion in every band. This is because the effective single-particle problem generated by our analysis gives rise to long-ranged hopping (see section \ref{sectiontwo}). Degeneracies in our band structures should therefore be taken seriously, despite the  simplicity of (\ref{Ham}).

The rest of the paper is organized as follows. In section \ref{sectiontwo}, we will review the graphical linked cluster technique and explain how the bands of the model (\ref{Ham}) are calculated. Section \ref{sectionthree} presents our results in the case of a magnetic field in the Kagome plane. The resulting phase is a magnonic Dirac semi-metal with Dirac points that can be manipulated by tuning the magnetic field or DMI. Section \ref{sectionfour} introduces an out-of-plane magnetic field, which breaks the symmetry protecting the Dirac points and gaps the spectrum. There we calculate the magnon thermal Hall conductivity, the Chern numbers of each band, and investigate a pair of topological phase transitions.

\section{Linked Cluster Expansion}\label{sectiontwo}

In this paper we employ the linked cluster expansion technique to derive the properties of single-magnon states. Here we will sketch the technique and establish notation; we refer the reader elsewhere for more details and proofs of our statements \cite{PhysRevLett.85.4373,2000AdPhy..49...93G,Gelfand1990,oitmaa_hamer_zheng_2006}. We begin by considering a model of spins on a lattice $\mathcal{L}$ with Hamiltonian

\begin{equation}
H = H_{0} + \lambda H_{1}
\end{equation}

where $H_{0}$ is a solvable Hamiltonian, $\lambda$ is a small parameter, and $H_{1}$ is a non-trivial perturbation. In our case, $1/|\vec{B}|$ plays the role of $\lambda$, and the unperturbed Hamiltonian is that of non-interacting spins in a magnetic field. The key to our analysis is to identify physical observables whose properties in the thermodynamic limit can be systematically understood by studying finite subsystems of $\mathcal{L}$. Define a connected cluster $c\subset\mathcal{L}$ to be an embedding of a connected graph into $\mathcal{L}$ (see figure \ref{fig:graphs}). The cluster $c$ inherits a cluster Hamiltonian $H_{c}$ by setting the couplings between all spins in $c$ and the remainder of the lattice to zero.

Let $P(\mathcal{L})$ be some physical property for the full lattice model, such as the ground state energy. Such a property can be computed on a cluster $c$, $P(c)$, using standard perturbation theory. For a generic physical property, computing $P(c)$ reveals little about $P(\mathcal{L})$. However, we will only consider properties which satisfy the following relation:

\begin{equation}\label{clusteraddition}
P(A+B) = P(A)\oplus P(B)
\end{equation}

This is the so-called cluster addition property, and it guarantees that we need only consider connected clusters. One proceeds by defining the weight of a cluster, $W(c)$,

\begin{equation}
W(c) = P(c) - \sum_{g\subset c}W(g)
\end{equation}

where $g$ indexes all (proper, connected) subclusters of $c$. In our case, the weight of a cluster with $n$ bonds will only contribute at order $\lambda^{n}$ due to the subgraph subtraction. Therefore, given a list of all connected graphs which have embeddings in $\mathcal{L}$ along with their multiplicities, we can compute $P(\mathcal{L})$ to arbitrarily high orders.

The discussion thus far works as described for simple properties such as the ground state energy. In general however we are interested in studying the excitations induced by $H_{1}$ about the ground state of $H_{0}$. For this purpose, the constraint (\ref{clusteraddition}) seems too strong: there is nothing preventing a quasiparticle from hopping between disconnected clusters. Moreover the number of excitations is not generally conserved unless it happens to be protected by a symmetry of $H$. Both problems are present in the model (\ref{Ham}), where the ground state of $H_{0}$ is spin-polarized and the excitations are magnons. We can circumvent the latter difficulty by generating effective Hamiltonians on subgraphs of $\mathcal{L}$ which forbid mixing between sectors with different quasiparticle numbers. This is done by finding a unitary transformation $U$ which block diagonalizes the cluster Hamiltonian $H_{c}$:

\begin{figure*}
  \includegraphics[width=1.0\linewidth]{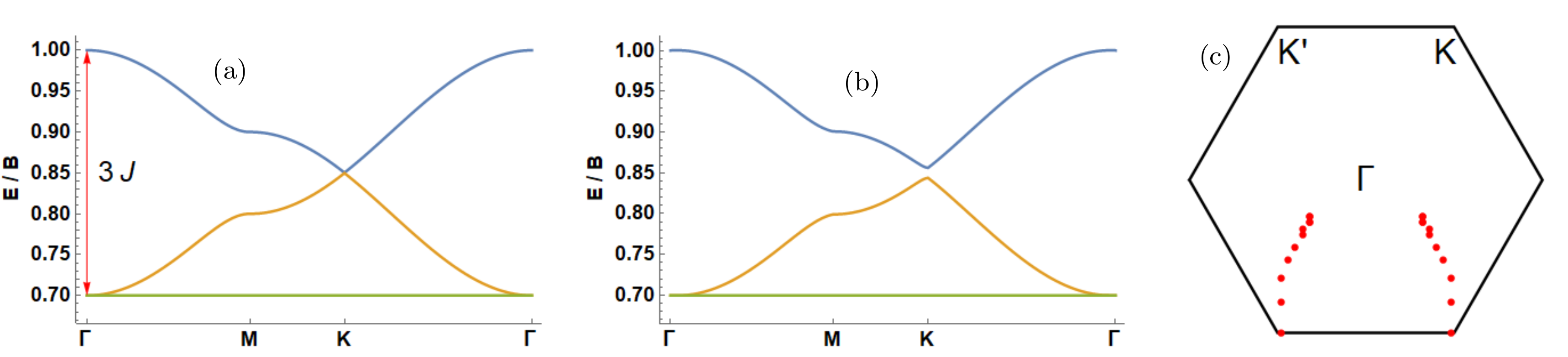}
  \caption{The band structure of (\ref{Ham}) for $\vec{B} = B\hat{x}, J = |\vec{B}|/10$ and various choices of DMI. (a): Without DMI, the band structure has Dirac points at $\vec{K},\vec{K'}$ and a dispersionless band. (b): Introducing out-of-plane DMI ($|\vec{D}| = J/10$) does not affect the Dirac points but the previously flat band becomes weakly dispersive (not visible). An in-plane component $(D_{p}=D_{z})$ of the DMI shifts the Dirac nodes off of high-symmetry points but does not gap the spectrum. (c): The spread of Dirac points as $\vec{D}$ is rotated into the plane. The nodes are displaced symmetrically about the magnetic field axis and orthogonal to it. These points are generated with ten uniformly spaced angular orientations of $\vec{D}$ in $\left[0,\pi/2\right]$. We set $|\vec{D}|=3J$ for visual effect.}
  \label{fig:bands}
\end{figure*}

\begin{equation}
H_{\text{eff}} = U^{\dagger}H_{c}U
\end{equation}

The transformation $U$ is constructed perturbatively in $\lambda$, so that $H_{\text{eff}}$ self-consistently forbids mixing up to a given order in perturbation theory. This allows us to study the single-particle band structure in spite of the lack of quasiparticle conservation.

The possibility of excitations hopping between disconnected clusters is still present. To avoid this we need to find a property related to the spectrum which satisfies (\ref{clusteraddition}). Define $E_{1}(\vec{i},\vec{j}) = \langle\vec{i}|H_{\text{eff}}|\vec{j}\rangle$ to be the hopping matrix element for a single particle state between sites $\vec{i}$ and $\vec{j}$. Then the following quantity has the cluster addition property:

\begin{equation}
\Delta_{1}(\vec{i},\vec{j}) = E_{1}(\vec{i},\vec{j}) - E_{0}\delta_{\vec{i},\vec{j}}
\end{equation}

where $E_{0}$ is the ground state energy of the cluster. We will only deal with systems invariant under translations by Bravais lattice vectors, which  encourages us to consider momentum eigenstates ($N$ is the number of lattice sites),

\begin{equation}
|\vec{k}\rangle = \frac{1}{\sqrt{N}}\sum_{\vec{j}}\exp\left(i\vec{k}\cdot\vec{j}\right)|j\rangle
\end{equation}

In general, we must allow for the possibility that the unit cells of $\mathcal{L}$ contain multiple sites. Let $\vec{\delta}_{ab}$ denote the vector connecting two sites of the lattice in sublattices $a$ and $b$ respectively (for the Kagome lattice, $a,b = 1,2,3$). Then the energetics of the quasiparticles are captured in the matrix

\begin{equation}\label{energetics}
\omega_{1}^{ab}(\vec{k}) = \sum_{\vec{\delta}_{ab}}\Delta_{1}(\delta_{ab})\left[\cos\left(\vec{k}\cdot\vec{\delta}_{ab}\right) + i\sin\left(\vec{k}\cdot\vec{\delta}_{ab}\right)\right]
\end{equation}

Diagonalizing this matrix gives the band structure at $\vec{k}$. This procedure generates effective tight binding Hamiltonians for one-particle excitations. In the case of the model (\ref{Ham}), we find that the matrix elements $E_{1}(\vec{i},\vec{j})$ are generically nonzero at sufficiently high orders in perturbation theory. This is a reflection of the strongly correlated nature of the magnons in this problem, and allows a model without NNN exchange to potentially capture details of the band structure of realistic systems.

\section{Tunable Dirac Points}\label{sectionthree}

In this section we take the magnetic field to lie in the Kagome plane, and unless otherwise mentioned we will let $\vec{B} = \hat{x}$ (this choice is not essential to the qualitative physics). In the absence of DMI and above the saturation field, the spectrum is known to contain a flat band at finite energy and two dispersive bands with Dirac points at $\vec{K} = \left(\frac{\pi}{3},\frac{\pi}{\sqrt{3}}\right), \vec{K'} = \left(-\frac{\pi}{3},\frac{\pi}{\sqrt{3}}\right)$ \cite{PhysRevB.82.184502, 2004PhRvB..70j0403Z, Owerre_2017} (figure \ref{fig:bands}). The tight binding model for the single magnon sector in this case involves only nearest-neighbor hopping.

Upon introducing DMI, the model develops dispersion in each band. We find that the hopping amplitudes satisfy

\begin{equation}\label{xsymmetry}
\Delta_{1}(\vec{\delta}_{ab}) = \Delta_{1}^{*}(-\vec{\delta}_{ab}) = \Delta_{1}\left(\vec{\delta}_{ba}\right)
\end{equation}

This result is consistent with a ground state ordering which is spin-polarized along an axis in the Kagome plane. Combined with (\ref{energetics}), this implies that $\omega_{1}^{ab}(\vec{k})$ is purely real. This result is manifest when the DMI points out of the Kagome plane, since the Hamiltonian is real; the extension to arbitrary DMI is less obvious. This implies that any effective two-band Hamiltonian obtained via projection has an expansion of the form

\begin{equation}\label{twoband}
H_{2\times 2}\left(\vec{k}\right) = E_{0}\mathbb{1} + h_{x}\left(\vec{k}\right)\sigma_{x} + h_{z}\left(\vec{k}\right)\sigma_{z}
\end{equation}

where $h_{x}(\vec{k}),h_{z}(\vec{k})$ are real-valued functions and $E_{0}>0$ is a constant energy shift. Perturbations which preserve (\ref{twoband}) are not expected to gap the Dirac points, which is consistent with our findings: tuning the relative magnitude of the in-plane ($D_{p}$) and out-of-plane ($D_{z}$) DMI shifts the Dirac points off of high-symmetry lines but never gaps them. The direction of displacement for the Dirac points also depends on the magnetic field orientation. This is to be contrasted with the typical result that SOC gaps out Dirac points, demonstrated for example in graphene.

\section{Topological Magnon Bands, Chern Numbers, and Thermal Hall Effect}\label{sectionfour}

\begin{figure}
  \includegraphics[width=0.85\linewidth]{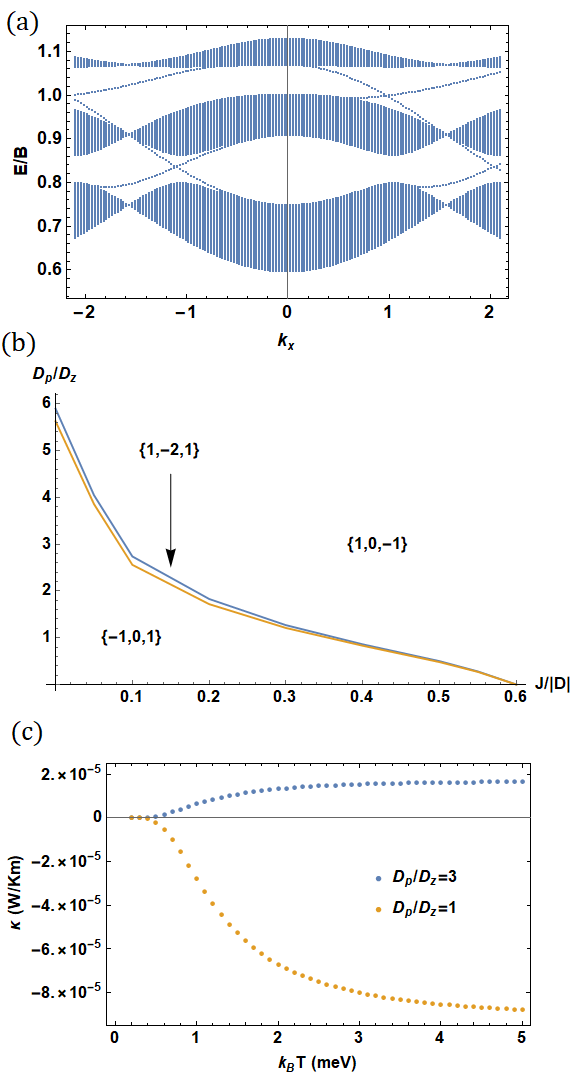}
  \caption{(a): Band structure for a semi-infinite system. The modes winding between the bulk bands are exponentially localized on the system's edges. (b): Phase diagram obtained by computing the band Chern numbers. We have chosen $J+|\vec{D}| = |\vec{B}|/5$. Different parameter choices only adjust the location of the phase boundary. The precise location of the phase boundaries is less significant than the necessary existence of small energy gaps near the transitions. The Chern numbers are indicated in each region from the highest to lowest energy bands; the set $\{1,-2,1\}$ describes the narrow region in parameter space indicated by the arrow.(c): Thermal Hall conductivity for $J/|\vec{D}| = 1/3$ both above and below the transition. We set the magnon energy scale to 5 meV in the absence of interactions, and $J+|\vec{D}| = 1$ meV. Hence $|\vec{D}| = 0.75$ meV and $J = 0.25$ meV here. The results are observed to be reasonably insensitive to this choice. The Hall conductivity changes sign across the transition by varying the ratio $D_{p}/D_{z}$ and the curve closer to the transition (yellow) is significantly enhanced due to the smallness of the gap.}
  \label{fig:thermalhall}
\end{figure}

With a magnetic field out of the Kagome plane, all lattice directions in the bulk are equivalent, and

\begin{equation}
\Delta_{1}(\vec{\delta}_{ab}) = \Delta_{1}(-\vec{\delta}_{ab}) = \Delta_{1}^{*}(\vec{\delta}_{ba})
\end{equation}

More generally, the hopping amplitudes are invariant under rotations which map the Kagome into itself, provided the sublattice structure is properly accounted for. $\omega_{1}^{ab}(\vec{k})$ is no longer manifestly real, which violates (\ref{twoband}). In two dimensions fine tuning is therefore required to find gapless points and a gapped spectrum is anticipated.

Previous work on similar systems has focused on the thermal Hall conductivity and we will do the same. By putting the system on a finite strip we find clear evidence of chiral edge states in the bulk band gaps which are associated with the conduction (figure \ref{fig:thermalhall} a). In the bulk, we compute the band Chern numbers \cite{doi:10.1143/JPSJ.74.1674}. Fixing $J+|\vec{D}|$ to be a constant fraction of $|\vec{B}|$, we can map out a phase diagram by looking for changes in the Chern numbers (figure \ref{fig:thermalhall} b). The most common set of Chern numbers is $\{1,0,-1\}$ (from the highest to lowest energy band). However there is also a phase with Chern numbers $\{-1,0,1\}$ separated from the former phase by a narrow region with Chern numbers $\{1,-2,1\}$. From an empirical perspective, these transitions are significant because they imply small energy gaps in the neighborhood of the phase boundaries. This typically enhances the thermal Hall conductivity, which we also find (figure \ref{fig:thermalhall} c). We also see that the thermal Hall conductivity changes sign across the phase boundary.

We compute the thermal Hall conductivity as follows. Letting $\Omega_{n}^{z}(\vec{k})$ denote the Berry curvature of band $n$, the (magnon) thermal Hall conductivity at temperature $T$ is given by \cite{PhysRevB.89.054420}

\begin{equation}\label{thermalhallcond}
\kappa_{xy} = -\frac{k_{B}^{2}T}{\left(2\pi\right)^{2}\hbar}\sum_{n}\int_{\text{BZ}}\left[c_{2}\left[g\left(\epsilon_{nk}\right)\right]-\frac{\pi^{2}}{3}\right]\Omega_{n}^{z}\left(\vec{k}\right)d^{2}k
\end{equation}

where $g(\epsilon_{nk})$ is the usual Bose-Einstein distribution factor. The function $c_{2}(x)$ is

\begin{equation}\label{c2}
c_{2}(x) = (1+x)\left[\ln\left(\frac{1+x}{x}\right)\right]^{2} - \left[\ln(x)\right]^{2}-2\text{Li}_{2}(-x)
\end{equation}

where $\text{Li}_{2}(x)$ is the dilogarithm. When stacks of Kagome layers are used with an interlayer spacing $\ell$, $\kappa/\ell$ is naturally given in W/Km. All values reported here assume $\ell = 5\AA$.

The values obtained for the thermal Hall conductivity are comparable to those found in other studies, although we have chosen small values of $J$ and $|\vec{D}|$ so that only modest magnetic fields are necessary to polarize the ground state. Specifically, figure \ref{fig:thermalhall} assumes a magnon energy of 5 meV in the absence of interactions, with $J+|\vec{D}| = 1$ meV.  The effect of stronger interactions on the thermal Hall conductivity can also be considered, but a price is paid by requiring a larger saturation field. In general, we see that there can be a significant benefit to finding materials which naturally sit near the phase boundaries discussed in figure \ref{fig:thermalhall}. The empirical interest in such a result should be clear to the spintronics community and others working with tunable DMI and other interactions.

\section{Conclusion and Future Directions}

In this paper, we have considered a ``minimal model'' of antiferromagnetic spins with SOC. In the absence of DMI, the physics of the magnons in the polarized phase is already well understood. The physics becomes significantly richer with the inclusion of DMI because of the coupling it induces between spin-space and real-space. By changing the ground state ordering (in our case, with a magnetic field), this coupling allows us to see qualitatively new physics, particularly more robust and controllable Dirac points.

We have also explored the phase diagram of a polarized magnet, and the presence of topological phase transitions opens up the possibility of finding enhanced response functions. A search for materials which can exhibit these enhanced responses could prove interesting.

As previously mentioned, our technique has the advantage of avoiding any linearization in the analysis. This means that multi-magnon states can be considered in detail with our technique, enabling the study of bound states and the multi-magnon continuum. This will be the subject of future work.

\section{Acknowledgements}

We would like to thank Eduardo da Silva Neto and Giacomo Resta for helpful discussions. This work is supported in part by NSF-DMR grant number 1855111.

\bibliography{KagomeAntiferromagnet_response}{}

\begin{thebibliography}{10}

\bibitem{LeonBalents2010Slif}
L.~Balents, ``Spin liquids in frustrated magnets,'' {\em Nature}, vol.~464,
  no.~7286, 2010.

\bibitem{Yan1173}
S.~Yan, D.~A. Huse, and S.~R. White, ``Spin-liquid ground state of the s = 1/2
  kagome heisenberg antiferromagnet,'' {\em Science}, vol.~332, no.~6034,
  pp.~1173--1176, 2011.

\bibitem{2015NatPh..11..444G}
P.~{Gegenwart} and S.~{Trebst}, ``{Spin-orbit physics: Kitaev matter},'' {\em
  Nature Physics}, vol.~11, pp.~444--445, June 2015.

\bibitem{PhysRevLett.64.88}
P.~Chandra, P.~Coleman, and A.~I. Larkin, ``Ising transition in frustrated
  heisenberg models,'' {\em Phys. Rev. Lett.}, vol.~64, pp.~88--91, Jan 1990.

\bibitem{1958JPCS....4..241D}
I.~{Dzyaloshinsky}, ``{A thermodynamic theory of ``weak'' ferromagnetism of
  antiferromagnetics},'' {\em Journal of Physics and Chemistry of Solids},
  vol.~4, pp.~241--255, 1958.

\bibitem{PhysRev.120.91}
T.~Moriya, ``Anisotropic superexchange interaction and weak ferromagnetism,''
  {\em Phys. Rev.}, vol.~120, pp.~91--98, Oct 1960.

\bibitem{2006AnPhy.321....2K}
A.~{Kitaev}, ``{Anyons in an exactly solved model and beyond},'' {\em Annals of
  Physics}, vol.~321, pp.~2--111, Jan 2006.

\bibitem{PhysRevLett.102.017205}
G.~Jackeli and G.~Khaliullin, ``Mott insulators in the strong spin-orbit
  coupling limit: From heisenberg to a quantum compass and kitaev models,''
  {\em Phys. Rev. Lett.}, vol.~102, p.~017205, Jan 2009.

\bibitem{2010NatPh...6..376P}
D.~{Pesin} and L.~{Balents}, ``{Mott physics and band topology in materials
  with strong spin-orbit interaction},'' {\em Nature Physics}, vol.~6,
  pp.~376--381, May 2010.

\bibitem{2013PhRvB..87f4423L}
P.~A. {Lee} and N.~{Nagaosa}, ``{Proposal to use neutron scattering to access
  scalar spin chirality fluctuations in kagome lattices},'' {\em \prb},
  vol.~87, p.~064423, Feb 2013.

\bibitem{Chen2013TailoringTC}
G.~Chen, T.~Ma, A.~T. N'Diaye, H.~Kwon, C.~Won, Y.~Wu, and A.~K. Schmid,
  ``Tailoring the chirality of magnetic domain walls by interface
  engineering.,'' {\em Nature communications}, vol.~4, p.~2671, 2013.

\bibitem{article}
G.~Chen and A.~K~Schmid, ``Imaging and tailoring the chirality of domain walls
  in magnetic films,'' {\em Advanced materials (Deerfield Beach, Fla.)},
  vol.~27, 05 2015.

\bibitem{2017RPPh...80a6502S}
L.~{Savary} and L.~{Balents}, ``{Quantum spin liquids: a review},'' {\em
  Reports on Progress in Physics}, vol.~80, p.~016502, Jan 2017.

\bibitem{2016NatCo...712691L}
F.-Y. {Li}, Y.-D. {Li}, Y.~B. {Kim}, L.~{Balents}, Y.~{Yu}, and G.~{Chen},
  ``{Weyl magnons in breathing pyrochlore antiferromagnets},'' {\em Nature
  Communications}, vol.~7, p.~12691, Sep 2016.

\bibitem{2018arXiv181101946O}
S.~A. {Owerre}, ``{Topological transitions of magnons in three-dimensional
  strained chiral antiferromagnets and thermal Hall effect in honeycomb
  ferromagnet CrI\$\_3\$},'' {\em arXiv e-prints}, p.~arXiv:1811.01946, Nov
  2018.

\bibitem{2018JPCM...30x5803O}
S.~A. {Owerre}, ``{Strain-induced topological magnon phase transitions:
  applications to kagome-lattice ferromagnets},'' {\em Journal of Physics
  Condensed Matter}, vol.~30, p.~245803, Jun 2018.

\bibitem{PhysRevB.66.014422}
M.~Elhajal, B.~Canals, and C.~Lacroix, ``Symmetry breaking due to
  dzyaloshinsky-moriya interactions in the kagom\'e lattice,'' {\em Phys. Rev.
  B}, vol.~66, p.~014422, Jul 2002.

\bibitem{PhysRevB.78.140405}
O.~C\'epas, C.~M. Fong, P.~W. Leung, and C.~Lhuillier, ``Quantum phase
  transition induced by dzyaloshinskii-moriya interactions in the kagome
  antiferromagnet,'' {\em Phys. Rev. B}, vol.~78, p.~140405, Oct 2008.

\bibitem{2015arXiv150801523S}
T.~F. {Seman}, C.-C. {Chen}, R.~R.~P. {Singh}, and M.~{van Veenendaal}, ``{The
  many faces of quantum kagome materials: Interplay of further-neighbour
  exchange and Dzyaloshinskii-Moriya interaction},'' {\em arXiv e-prints},
  p.~arXiv:1508.01523, Aug 2015.

\bibitem{PhysRevLett.98.207204}
M.~Rigol and R.~R.~P. Singh, ``Magnetic susceptibility of the kagome
  antiferromagnet ${\mathrm{zncu}}_{3}(\mathrm{OH}{)}_{6}{\mathrm{cl}}_{2}$,''
  {\em Phys. Rev. Lett.}, vol.~98, p.~207204, May 2007.

\bibitem{PhysRevB.95.054418}
M.~Hering and J.~Reuther, ``Functional renormalization group analysis of
  dzyaloshinsky-moriya and heisenberg spin interactions on the kagome
  lattice,'' {\em Phys. Rev. B}, vol.~95, p.~054418, Feb 2017.

\bibitem{PhysRevB.85.104417}
V.~A. Zyuzin and G.~A. Fiete, ``Spatially anisotropic kagome antiferromagnet
  with dzyaloshinskii-moriya interaction,'' {\em Phys. Rev. B}, vol.~85,
  p.~104417, Mar 2012.

\bibitem{PhysRevLett.117.187203}
A.~L. Chernyshev and P.~A. Maksimov, ``Damped topological magnons in the
  kagome-lattice ferromagnets,'' {\em Phys. Rev. Lett.}, vol.~117, p.~187203,
  Oct 2016.

\bibitem{PhysRevB.98.224414}
C.-Y. Lee, B.~Normand, and Y.-J. Kao, ``Gapless spin liquid in the kagome
  heisenberg antiferromagnet with dzyaloshinskii-moriya interactions,'' {\em
  Phys. Rev. B}, vol.~98, p.~224414, Dec 2018.

\bibitem{Owerre_2016}
S.~A. Owerre, ``Magnon hall effect without dzyaloshinskii{\textendash}moriya
  interaction,'' {\em Journal of Physics: Condensed Matter}, vol.~29,
  p.~03LT01, nov 2016.

\bibitem{PhysRevLett.104.066403}
H.~Katsura, N.~Nagaosa, and P.~A. Lee, ``Theory of the thermal hall effect in
  quantum magnets,'' {\em Phys. Rev. Lett.}, vol.~104, p.~066403, Feb 2010.

\bibitem{2018PhRvB..97h1106R}
A.~{R{\"u}ckriegel}, A.~{Brataas}, and R.~A. {Duine}, ``{Bulk and edge spin
  transport in topological magnon insulators},'' {\em \prb}, vol.~97,
  p.~081106, Feb 2018.

\bibitem{2019arXiv190105683C}
W.~{Cai}, J.~{Han}, F.~{Mei}, Y.~{Xu}, Y.~{Ma}, X.~{Li}, H.~{Wang}, Y.~{Song},
  Z.-Y. {Xue}, Z.-q. {Yin}, S.~{Jia}, and L.~{Sun}, ``{Observation of
  topological magnon insulator states in a superconducting circuit},'' {\em
  arXiv e-prints}, p.~arXiv:1901.05683, Jan 2019.

\bibitem{2019arXiv190504549M}
F.~{Mei}, G.~{Chen}, N.~{Goldman}, L.~{Xiao}, and S.~{Jia}, ``{Topological
  magnon insulator and quantized pumps from strongly-interacting bosons in
  optical superlattices},'' {\em arXiv e-prints}, p.~arXiv:1905.04549, May
  2019.

\bibitem{PhysRevLett.82.4536}
M.~E. Zhitomirsky and A.~L. Chernyshev, ``Instability of antiferromagnetic
  magnons in strong fields,'' {\em Phys. Rev. Lett.}, vol.~82, pp.~4536--4539,
  May 1999.

\bibitem{PhysRevB.89.134409}
A.~Mook, J.~Henk, and I.~Mertig, ``Magnon hall effect and topology in kagome
  lattices: A theoretical investigation,'' {\em Phys. Rev. B}, vol.~89,
  p.~134409, Apr 2014.

\bibitem{2018PhRvB..98i4419L}
P.~{Laurell} and G.~A. {Fiete}, ``{Magnon thermal Hall effect in kagome
  antiferromagnets with Dzyaloshinskii-Moriya interactions},'' {\em Physical
  Review B}, vol.~98, p.~094419, Sep 2018.

\bibitem{PhysRevB.73.214446}
T.~Yildirim and A.~B. Harris, ``Magnetic structure and spin waves in the
  kagom\'e jarosite compound
  $\mathrm{K}{\mathrm{fe}}_{3}{(\mathrm{S}{\mathrm{O}}_{4})}_{2}{(\mathrm{O}\mathrm{H})}_{6}$,''
  {\em Phys. Rev. B}, vol.~73, p.~214446, Jun 2006.

\bibitem{PhysRevLett.96.247201}
K.~Matan, D.~Grohol, D.~G. Nocera, T.~Yildirim, A.~B. Harris, S.~H. Lee, S.~E.
  Nagler, and Y.~S. Lee, ``Spin waves in the frustrated kagom\'e lattice
  antiferromagnet
  ${\mathrm{kfe}}_{3}(\mathrm{OH}{)}_{6}({\mathrm{so}}_{4}{)}_{2}$,'' {\em
  Phys. Rev. Lett.}, vol.~96, p.~247201, Jun 2006.

\bibitem{PhysRevB.92.094409}
A.~L. Chernyshev, ``Strong quantum effects in an almost classical
  antiferromagnet on a kagome lattice,'' {\em Phys. Rev. B}, vol.~92,
  p.~094409, Sep 2015.

\bibitem{PhysRevLett.85.4373}
S.~Trebst, H.~Monien, C.~J. Hamer, Z.~Weihong, and R.~R.~P. Singh,
  ``Strong-coupling expansions for multiparticle excitations: Continuum and
  bound states,'' {\em Phys. Rev. Lett.}, vol.~85, pp.~4373--4376, Nov 2000.

\bibitem{2000AdPhy..49...93G}
M.~P. {Gelfand} and R.~R.~P. {Singh}, ``{High-order convergent expansions for
  quantum many particle systems},'' {\em Advances in Physics}, vol.~49,
  pp.~93--140, Jan. 2000.

\bibitem{Gelfand1990}
M.~P. Gelfand, R.~R.~P. Singh, and D.~A. Huse, ``Perturbation expansions for
  quantum many-body systems,'' {\em Journal of Statistical Physics}, vol.~59,
  pp.~1093--1142, Jun 1990.

\bibitem{oitmaa_hamer_zheng_2006}
J.~Oitmaa, C.~Hamer, and W.~Zheng, {\em Series Expansion Methods for Strongly
  Interacting Lattice Models}.
\newblock Cambridge University Press, 2006.

\bibitem{PhysRevB.82.184502}
S.~D. Huber and E.~Altman, ``Bose condensation in flat bands,'' {\em Phys. Rev.
  B}, vol.~82, p.~184502, Nov 2010.

\bibitem{2004PhRvB..70j0403Z}
M.~E. {Zhitomirsky} and H.~{Tsunetsugu}, ``{Exact low-temperature behavior of a
  kagom{\'e} antiferromagnet at high fields},'' {\em Physical Review B},
  vol.~70, p.~100403, Sep 2004.

\bibitem{Owerre_2017}
S.~A. Owerre, ``Magnonic analogs of topological dirac semimetals,'' {\em
  Journal of Physics Communications}, vol.~1, p.~025007, sep 2017.

\bibitem{doi:10.1143/JPSJ.74.1674}
T.~Fukui, Y.~Hatsugai, and H.~Suzuki, ``Chern numbers in discretized brillouin
  zone: Efficient method of computing (spin) hall conductances,'' {\em Journal
  of the Physical Society of Japan}, vol.~74, no.~6, pp.~1674--1677, 2005.

\bibitem{PhysRevB.89.054420}
R.~Matsumoto, R.~Shindou, and S.~Murakami, ``Thermal hall effect of magnons in
  magnets with dipolar interaction,'' {\em Phys. Rev. B}, vol.~89, p.~054420,
  Feb 2014.

\end{thebibliography}

\end{document}